\documentclass[%
 aip,
 amsmath,amssymb,
 reprint,%
]{revtex4-1}
\usepackage[english]{babel}
\usepackage{hyperref}
\usepackage{cleveref}
\usepackage{graphicx}
\usepackage{dcolumn}
\usepackage{bm}
\usepackage{natbib}
\usepackage[utf8]{inputenc}
\usepackage[T1]{fontenc}
\usepackage{mathptmx}
\usepackage{etoolbox}
\usepackage{xcolor}
\usepackage{float}
\usepackage[multiple]{footmisc}

\newcommand{\angstrom}{\textup{\AA}}
\newcommand{\reference}{\textnormal{ref.}}
\newcommand{\simulation}{\textnormal{sim.}}

\begin{document}


\title{Learning Pair Potentials using Differentiable Simulations} 



\author{Wujie Wang} 
\thanks{equal contribution}
\affiliation{Department of Materials Science and Engineering, Massachusetts Institute of Technology. 77 Massachusetts Ave. Cambridge MA 02139 USA}

\author{Zhenghao Wu}
\thanks{equal contribution}
\affiliation{Eduard-Zintl-Institut f\"{u}r Anorganische und Physikalische Chemie, Technische Universit\"{a}t Darmstadt, Alarich-Weiss-Str. 8, 64287 Darmstadt, Germany}

\author{Rafael G\'omez-Bombarelli}
\email[]{rafagb@mit.edu}
\affiliation{Department of Materials Science and Engineering, Massachusetts Institute of Technology. 77 Massachusetts Ave. Cambridge MA 02139 USA}

\date{\today}

\begin{abstract}
Learning pair interactions from experimental or simulation data is of great interest for molecular simulations. We propose a general stochastic method for learning pair interactions from data using differentiable simulations (DiffSim). DiffSim defines a loss function based on structural observables, such as the radial distribution function, through molecular dynamics (MD) simulations. The interaction potentials are then learned directly by stochastic gradient descent, using backpropagation to calculate the gradient of the structural loss metric with respect to the interaction potential through the MD simulation. This gradient-based method is flexible and can be configured to simulate and optimize multiple systems simultaneously. For example, it is possible to simultaneously learn potentials for different temperatures or for different compositions. We demonstrate the approach by recovering simple pair potentials, such as Lennard-Jones systems, from radial distribution functions. We find that DiffSim can be used to probe a wider functional space of pair potentials compared to traditional methods like Iterative Boltzmann Inversion. We show that our methods can be used to simultaneously fit potentials for simulations at different compositions and temperatures to improve the transferability of the learned potentials.

\end{abstract}


\maketitle 

\section{Introduction}

Molecular simulation has been an invaluable technique in various fields of modern science, including chemistry, physics, and materials science\cite{vlachakis_current_2014,blumberger_recent_2015,gartner_modeling_2019,fish_mesoscopic_2021}. Ideally, an atomistic computer simulation would combine the accurate description of the Born-Oppenheimer potential energy surface (PES) with the treatment of nuclear motion, both at the quantum mechanics level. However, this is computationally infeasible for most of the relevant molecular sizes and time scales. It is typically necessary to approximate the PES with simpler and faster surrogate functions, such as force fields or interatomic potentials. 
Therefore, the development of accurate and computationally efficient force fields has become one of the most important topics in molecular simulations\cite{mayo_dreiding_1990,jorgensen_development_1996,pastor_development_2011}. 

Covalent bonds in organic molecules are typically described in force fields as harmonic or polynomial expressions in short-range many-body terms (bonds, angles, torsions), while numerous types of description have been proposed for noncovalent interactions\cite{rapaport_art_2004,cisneros_modeling_2016}. Of all these variants, the pairwise potential is most commonly used in molecular simulations, as it has been shown to be sufficient to capture significant molecular mechanics, and is also the most computationally cost-effective way to describe noncovalent interactions between particles\cite{cisneros_modeling_2016}. For example, rigid water models with pairwise additive PESs such as SPC/E\cite{berendsen_missing_1987} and TIP4P\cite{jorgensen_comparison_1983} recover many thermodynamic properties, e.g., densities and heat capacities, that are consistent with experimental measurements across a wide range of temperatures. These models with pairwise additive PESs are efficient in terms of computational cost and scalability\cite{glaser_strong_2015}. It is noted that the many-body expansion of PESs has been widely appreciated for its role in accurately predicting behaviors or properties, such as phase transitions of water, which the pairwise additive PES cannot predict\cite{cisneros_modeling_2016,albaugh_advanced_2016}. However, these potential formulations that carry the many-body effect are usually not as efficient; depending on the size of the system, they are usually 10 to 1000 times slower to simulate than pairwise potentials.\cite{plimpton_computational_2012}

In addition to all-atom potentials, the derivation of effective pair potentials also plays an important role for the efficient simulation of coarse-grained systems.\cite{muller-plathe_coarse-graining_2002,souza_martini_2021,dhamankar_chemically_2021} Various variational frameworks have been developed to derive coarse-grained potentials. Based on the idea of force matching\cite{ercolessi_interatomic_1994}, Izvekov and Voth developed a force-based coarse-graining method, i.e., multiscale coarse-grained method (MS-CG)\cite{izvekov_multiscale_2005}. The MS-CG method aims to reproduce the many-body potential of the mean force (PMF) of the fine-grained system (FG) in the mapped CG representation by matching the force acting on the CG bead.\cite{noid_multiscale_2007} This framework can also be extended to different architectures of force fields, e.g., machine learning potentials, optimized with stochastic gradient descent. \cite{zhang2018deepcg, wang_machine_2019}. However, it is a known issue that the potentials learned from force matching do not necessarily recover target structural distributions due to missing cross-correlations between degrees of freedom\cite{rudzinski2012role} The use of machine learning potentials naturally incorporates many-body correlation terms, but they often suffer from instabilities in dynamics and often fail to generate robust simulation trajectories. \cite{wang_machine_2019, stocker2022robust}. Beyond force matching, Shell et al. developed a series of coarse-graining approaches using relative entropy to measure coarse-graining information loss\cite{chaimovich_coarse-graining_2011}. The standard relative entropy framework uses the Newton-Raphson algorithm as the solver, which, however, is computationally expensive for optimizing models such as deep neural networks that have a large number of parameters. \cite{kocer_neural_2022} There is another class of structure-based coarse-graining method which does not require atomistic position or force data and directly aims at reproducing structural distributions, of which iterative Boltzmann inversion (IBI) is the most notable.\cite{reith_deriving_2003} IBI constructs potentials for a CG model by iteratively correcting errors in simulated radial distribution function until convergence.\cite{reith_deriving_2003} It has been widely employed for coarse-graining various soft-matter systems such as organic liquids\cite{qian_temperature-transferable_2008}, ionic liquids\cite{karimi-varzaneh_studying_2010} and polymers\cite{qian_temperature-transferable_2008}. IBI does not necessarily require reference all-atom simulations. However, its application is limited to tabulated pair potentials and does not generalize to other forms of potential.


Most coarse-graining approaches only aim at reproducing statistics in a single thermodynamic state, so the developed potentials are rarely transferable. Several approaches have been proposed to improve the transferability of the coarse-graining frameworks mentioned in previous sections. For example, Mullinax and Noid proposed an extended ensemble method in the MS-CG framework\cite{mullinax_extended_2009}, in which datasets from multiple equilibrium ensembles are collected to develop CG potentials that reproduce the potential of mean force of FG systems in these ensembles. This approach requires a considerable amount of atomistic force data, which hinders its application.

CG models created by the IBI method can reproduce the structural correlations of the FG model at a single thermodynamic state, but they do not produce potentials that are transferable to a wide range of thermodynamic states with different temperatures and compositions. Attempts have been made to derive transferable CG potentials using multi-state data based on IBI, but these are still challenging, for example, in finding appropriate weighting functions.\cite{moore_derivation_2014} 


In recent years, machine learning (ML) has emerged as an important tool for molecular modeling. Much of the success has been based on the application of differentiable programming, which enables the exact generation of gradient programs. 
The process of generating the gradient program is termed Automatic Differentiation (AD). Although emerging as a relatively new tool, AD has shown great progress in many domains of computational science, such as quantum computing\cite{liao2019differentiable} and fluid dynamics\cite{schenck2018spnets}. For the modeling of PES, AD has transformed the fitting of force fields from first-principle calculations: once a forward program that transforms nuclear coordinates to a scalar (energy) is constructed, the gradient (force) program is automatically generated with AD. The generated force field is also trainable with respect to the data, enabling parameterization of ML potentials with near ab initio accuracy.\cite{unke2021machine}

Differentiable programming is applicable to many controlled iterative mathematical procedures in the physical sciences, ranging from learning exchange correlation functionals for Density Functional Theory(DFT) calculations from data\cite{kasim2021learning, kanungo2019exact} or fine-tuning molecular basis sets\cite{tamayo2018automatic}. This requires differentiating through mathematical procedures such as fixed-point iteration \cite{blondel2021efficient} and eigendecomposition\cite{magnus1985differentiating}. Similarly, differentiation operations of molecular simulations, as solutions to some ordinary differential equation (ODE), can also be defined.\cite{chen2018neural} Recently, many applications have been proposed using differentiable molecular dynamics (DiffSim). \cite{wang2020differentiable, schoenholz2020jax, doerr2021torchmd, thaler2021learning, greener2021differentiable} Specifically, recent work on DiffSim suggests a differentiable top-down approach to learning interatomic potentials directly based on macroscopic observations from experiments or simulations. Compared to previously proposed machine learning potentials that are trained from the bottom-up, DiffSim enables experimentally informed parameterization of force fields. However, to improve fitting flexibility, these methods utilize neural network (NN) potentials based on graph neural networks, which have limited interpretability and are less scalable for large-scale simulations. Another method of machine learning-assisted top-down parameterization involves directly predicting force field parameters from macroscopic structure correlation using a machine-learned function to bypass the inverse problem.\cite{moradzadeh2019transfer} This approach requires abundant labeled simulation data and is limited to a simple functional form of Lennard-Jones type, which has limited capacity to encode complex structural correlations. 

In this work, we propose simulation-informed parameterization of pair potentials using DiffSim. Compared to structure-based methods like IBI, our approach can be used to optimize flexible functional forms and can be easily configured to fit multiple systems simultaneously; compared to force-matching-based methods, our method does not require atomistic position or force data and aims at directly reproducing the target distributions. First, we introduce our method and training protocols.  For numerical examples, we demonstrate our method on simple liquid systems where ground truth potentials are known. We show that our method is good at obtaining a diverse set of pair potentials that reproduce the target liquid structure. To demonstrate the flexibility of the method, we show that the model can learn pair potentials for binary mixture systems involving multiple interactions and compositions. Our results indicate that fitting multiple states simultaneously helps to improve the transferability of learned potentials. Furthermore, our method can be used to learn coarse-grained potentials where there is no ground truth for pairwise potentials. We show that the learning of interactions can be combined with flexible functional forms that involve temperature dependence for improved transferability.

\section{Method}
\subsection{Observable-based Differentiable Simulations}

\begin{figure*}
    \centering
    \includegraphics[width=0.9\textwidth]{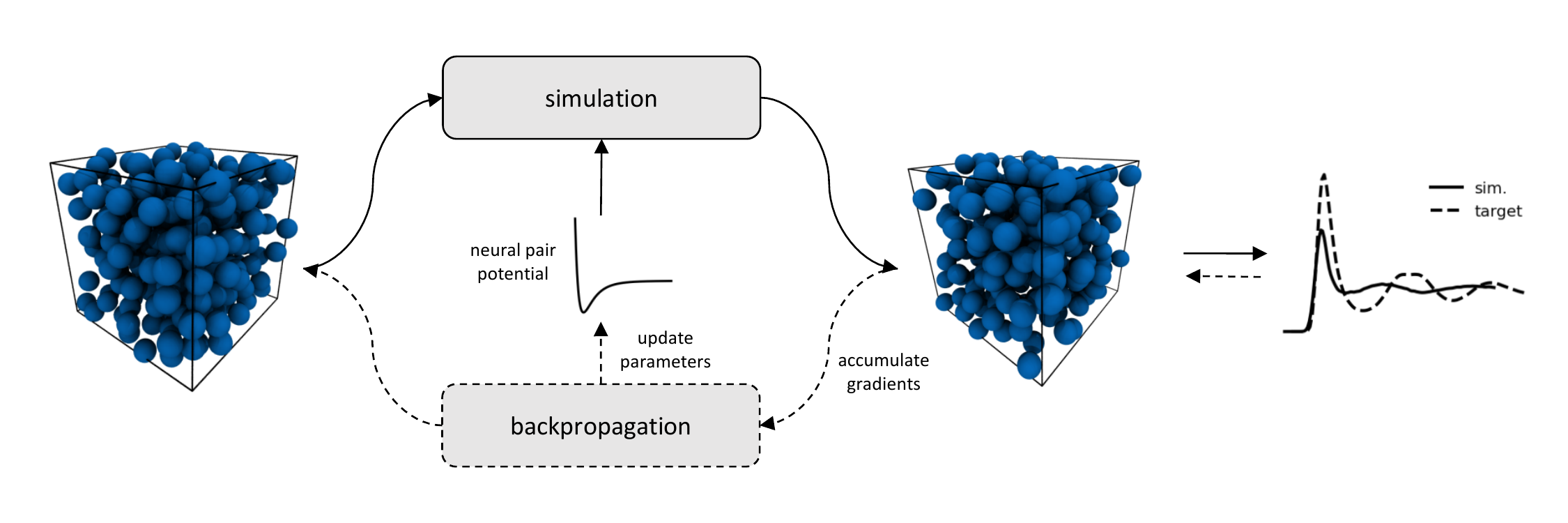}
    \caption{A schematic diagram for our learning framework}
    \label{fig:schematic}
\end{figure*}

Differentiable programming is a programming paradigm in which the gradient of a mathematical procedure can be automatically synthesized. In deep learning, automatically computed gradients allow gradient-based optimization using ``backpropagation" to train deep neural networks. As universal approximators, neural networks of various architectures have been used as a drop-in module for optimization and learning. However, without the appropriate incorporation of inductive biases, many NN models tend to overfit the data and often fail to generalize.\cite{battaglia2018relational} With the development of differentiable algorithms, many sampling and optimization procedures can now be directly incorporated into the machine learning framework. These structures serve as inductive biases to regularize learning, improve data efficiency, and improve model interpretability. \cite{baxter2000model} Machine learning of dynamical systems and differentiable equations is a domain where differentiable sampling operations are used as strong inductive biases. \cite{han2018solving, chen2017tutorial, brunton2022data}

In molecular dynamics, the simulation of systems of many particles requires integrating per-particle state variables in time following the governing equations of motion. Although continuous, the computational solution of state variable dynamics is solved in discrete time steps, involving the composition of differentiable mathematical procedures. With AD, the computation of observables from molecular simulations is, therefore, end-to-end differentiable. Specifically, given a defined forward computational program, the AD framework traces the computation backward and recursively applies chain rules as vector Jacobian products. This requires the implementation of the accompanying gradient function for every primitive operation used for the forward computation. AD can be applied to arbitrarily complex numerical programs, as long as the gradient of each elementary step can be computed. Auto-differentiable code has been implemented in many modern numerical computing packages. \cite{abadi2016tensorflow, frostig2018compiling, paszke2019pytorch}

 The evolution of atomic coordinates in MD is typically obtained by integrating the equations of motion through a discretized update rule of the following form: 
\begin{equation}
    x_{t} = F_{\theta}(x_{t-1}, v_{t-1}, \eta_{t-1}) dt + x_{t-1}
\end{equation}

where $x$ is the coordinate, $v$ is the velocity, $\eta$ is the bath variable, $\theta$ represents the set of learning parameters, and $F$ is the discretized update procedure. The simulation observables $O$ are usually extracted from the trajectories. Examples of such observables include radial distribution functions (RDF), velocity autocorrelation functions, etc. To learn or control the simulated observables, a scalar loss function $L$ must be defined to quantify the discrepancy between the simulated and target observables. The loss function $L$ needs to be minimized using gradient descent with gradient on $\theta$. We first consider the computation of the gradient on $x$, or the adjoint state, at each time step:

\begin{equation}
\begin{aligned}
    \frac{dL}{dx_t} &= \frac{\partial L}{ \partial x_t } + \frac{dL}{dx_{t+1}} \frac{\partial x_{t+1}}{\partial x_t}\\
        &=\frac{\partial L}{ \partial x_t }  + \frac{dL}{dx_{t+1}} + \frac{dL}{dx_{t+1}} \frac{\partial F_{\theta} }{\partial x_t} dt
\end{aligned}
\label{eq:grad}
\end{equation}

With $\frac{dL}{d xt_t}$ at each step $t$, the gradient of $\theta$ can be accumulated using the chain rule:

\begin{equation}
    \frac{dL}{d\theta} =  \sum_t^{T} \frac{dL}{dx_t}\frac{dx_t}{d\theta}= \sum_t^{T} \frac{dL}{dx_t}\frac{dF_{\theta} (x_{t-1}, ...)}{d\theta} dt
    \label{eq:adjoint}
\end{equation}

Alternatively, one can treat \cref{eq:adjoint} as a dynamical system solved backward in time.\cite{chen2018neural} With the gradient accumulated from MD simulations for $T$ steps, gradient-based optimization can be applied to minimize the observable optimization objective $L$ until convergence.

\subsubsection{Differentiable Computation of Radial Distributions Functions}
The goal of our method is to learn pair interactions from RDFs, which are treated as our observable $O$. RDF is defined as: 
    \begin{equation}
        g(r) = \frac{V}{N^2 4 \pi r^2} p(r)
    \end{equation}

where $V$ is the volume, $N$ is the number of particles and $r$ is the radial distance from a target particle; $p(r)$ is the histogram of pair distances. To propagate the gradient from $O$ to the model parameters, $p(r)$ must be differentiable, while the operation to obtain a histogram is not differentiable. To make the computation of $g(r)$ differentiable, we apply the kernel density trick with Gaussian kernels to make the procedure for computing histograms differentiable. \cite{chen2017tutorial} By definition, 
    \begin{equation}
        p(r) = \sum_{i \neq j} \frac{\delta(r_{ij} - r)}{\sum_{i \neq j} 1}
    \end{equation}
    
where $r_{ij}$ is the distance between the particle $i$ and $j$, and $\delta (\cdot)$ is the Dirac Delta function. To estimate $\delta(\cdot)$ in a differentiable way, we approximate the probability density of two particles that are $\mu_k$ apart in distance as:
\begin{equation}
    p(\mu_k) \approx \sum_{i \neq j} p_k(r_{ij}) = \sum_{i \neq j} \frac{e^{-(\mu_k - r_{ij})^2/d}}{\sum_{k'=0}^{k=K-1} e^{-(\mu_{k'} - r_{ij})^2/d}}
\end{equation}
    
where $d = \mu_{k+1} - \mu_k$ is the size of the histogram bin and $K$ is the total number of evenly spaced Gaussian kernels. The summation in the denominator ensures that the histogram is normalized over $[\mu_0, \mu_{K-1}]$. This approximation provides a density estimate over binned domains. Similar methods with triangular kernels have also been proposed before.\cite{ustinova2016learning} As $d \rightarrow 0$, the Gaussian kernel function approximates $\delta(\cdot)$.

\subsubsection{Design for Neural Pair Potentials}

We propose a design for the neural network $NN:  \mathbb{R} \rightarrow \mathbb{R}$ with a parameter set $\theta$ as an approximation function for pair potentials. We choose neural networks (NN) to parameterize pair potentials, because NNs are universal approximators\cite{hornik1989multilayer} and can be easily stacked in a model for end-to-end training. The approach is also applicable to parameterizing fixed functional forms such as Lennard-Jones or other choices. Given a distance $r_ij$ within a cutoff distance $r_{\mathrm{cut}}$, we expand the distance with a Gaussian basis to obtain a distance feature:
        \begin{equation}
            e_k(r_{ij}) = e^{- \frac{(\mu_k - r_{ij})^2}{d_k}}
        \end{equation}
        
where the initial $d_k $ is learnable with the initial value set as $r_{\mathrm{cut}} / K$ and $K$ is the number of Gaussian bases used to parameterize $r_{ij}$. The $K$ dimensional distance feature is then fed into a standard multilayer perceptron (MLP) with $n_{\mathrm{layer}}$ hidden layers and $W$ hidden nodes per layer to parameterize a scalar energy output. We term our neural pair potential $u_\theta$ with $\theta$ indicating the set of learnable parameters that include $d_k$ and weights and biases in MLP. AD can be applied to generate the force on an individual particle $i$ to be used for simulation. The force program is also differentiable to receive gradient signals to adjust the parameter set $\theta$. In addition to the MLP-based pair potential, we also include an unlearnable prior pair potential $U_{\mathrm{prior}}$ to guide the initial simulations to explore reasonable regions of the configuration space. For $u_{\mathrm{prior}}$, we simply use a repulsive potential of the form $\sum_{i, j}(\frac{\sigma}{r_{ij}})^c$ where $c$ is the repulsive exponent, which is also treated as a hyperparameter. The total pair potentials used for DiffSim take the form:
\begin{equation}
    U_{\theta}(x) = \sum_m \sum_{i, j \in \mathcal{P}_m} u^m_\theta(r_{ij}) + u_{prior}(r_{ij})
\end{equation}

where $m$ indicates the type of pair interactions and $\mathcal{P}_m$ indicates the set of atom pairs that are governed by the pair potential of the same type.  In case faster convergence is needed, we also perform an extra pre-training step to set the pair potentials with the direct Boltzmann inversion\cite{tschop1998simulation} so that $u^m_\theta(r_{ij}) + u_{prior}(r_{ij}) \approx -kT \ln g^m(r)$ with $g^m(r)$ being the target RDF.

\subsubsection{Learning Protocols}

In this section, we introduce our protocols for learning neural pair potentials. The learning protocol follows the concept of learning-to-simulate,\cite{ruiz2018learning} with the simulator performing the coupled task of sampling data points and learning.  We first initialize the MD systems as crystal structures (FCCs or BCCs) at a target density. We then apply DiffSim to evolve the system for $T$ steps and compute the observable of interests. The examples used in this work are simulated in the canonical ensemble (NVT) using the Nose-Hoover chain integrator.\cite{martyna1992nose} To learn pair potentials, we are interested in learning a $U_{\theta}$ that produces a target RDF $g_{\reference}(r)$. We term the RDF obtained from DiffSim as $g_{\simulation}(r)$. We then want to minimize the L2 loss between targets and simulated RDFs:
    \begin{equation}
        L = \sum_m \int_r (g^m_{\simulation}(r) - g^m_{\reference}(r))^2  dr 
    \label{eq:loss}
    \end{equation}

where $m$ again indicates the type of pair potential. With DiffSim, we can then obtain gradients on the parameters according to \cref{eq:adjoint}. The simulation explores the configuration space with an initial potential and accumulates gradients based on the simulated trajectories. 

Gradients can be fed into a gradient-based optimizer, such as Adam\cite{kingma2014adam}, to update $U_{\theta}$. We simulate and update the pair potentials for $N_{epoch}$ epochs until the learning converges. Because differentiable simulations can be susceptible to exploding or vanishing gradients \cite{ingraham2018learning, metz2021gradients}, we tend not to evolve the system too long before each gradient updates. In practice, we observe poor learning outcomes with simulation steps that are more than 300 steps. We applied small learning rates to perturb the simulation slightly between forward and backward passes. The learning rate is also scheduled to decrease if it reaches a plateau during learning. The decreasing learning rate facilitates fine-tuning of the learned potentials when the sampled configurations produce an observation close to the target. To report the equilibrium RDF, we simulate our example systems sufficiently long to reflect the true RDF governed by the learned pair interactions. In our experiments, we choose $T$ between 100 and 300 to ensure a stable gradient calculation and to provide sufficient time for the system to evolve based on the updated force field. 

\subsection{Iterative Boltzmann Inversion}

We compared our proposed learning protocol with iterative Boltzmann inversion (IBI), a popular structure-based method to determine effective pair potentials from RDF. \cite{reith_deriving_2003} IBI is the iterative version of the Boltzmann inversion (BI).\cite{tschop1998simulation} BI is a simple method based on the fact that the distribution of an independent degree of freedom $x$ corresponds to a Boltzmann distribution in a canonical ensemble:
\begin{equation}
    p(x)\propto \exp(-\frac{u(x)}{k_BT})
\end{equation}
where $p(x)$ is the normalized distribution; $k_B$ and $T$ are the Boltzmann constant and the temperature, respectively. The potential $u(x)$ can be obtained by inverting the above equation:
\begin{equation}
    u(x)= -k_BT\log g_{\reference}(x)
\end{equation}
The potential obtained is in the form of tabulated pair potentials. Such a direct inversion ignores the indirect force contribution to the pair statistics, and therefore introduces errors in reproducing the target RDF. \cite{tschop1998simulation, reith_deriving_2003} As a natural extension of BI, IBI iteratively corrects the potential using the difference between the target distribution and the distribution obtained from the trial potential energy function: 
\begin{equation}
    u_{t+1}(x)=u_{t}(x)+\alpha \Delta u_t
\end{equation}
and
\begin{equation}
    \alpha \Delta u_t = k_BT\ln\big(\frac{g_{t}(x)}{g_{\mathrm{\reference}}(x)}\big)
\end{equation}

where $u_{t}$ is the effective potential at iteration $t$ and $\alpha$ is the size of the update step. Here, $g_{t}(x)$ is the RDF simulated with $u_t$, and $g_{\mathrm{\reference}}(x)$ represents the RDF calculated from a reference system. Since the IBI is only compatible with tabulated pair potentials, it cannot be used to optimize a broader class of potentials such as neural networks and pair potentials with explicit thermodynamic dependence, which are used in this work. Practically, IBI also suffers from convergence issues, as in some cases hundreds of iterations are required to converge. \cite{potestio2013henderson,rosenberger_comparison_2016} In our work, we use IBI as a baseline comparison, and all CG potentials are obtained using the Versatile Object-Oriented Toolkit for Coarse-Graining Applications (VOTCA). \cite{ruhle_versatile_2009}

\section{Results}

\subsection{Learning Simple Pair Potentials}

\begin{figure}
    \centering
    \includegraphics[width=0.5\textwidth]{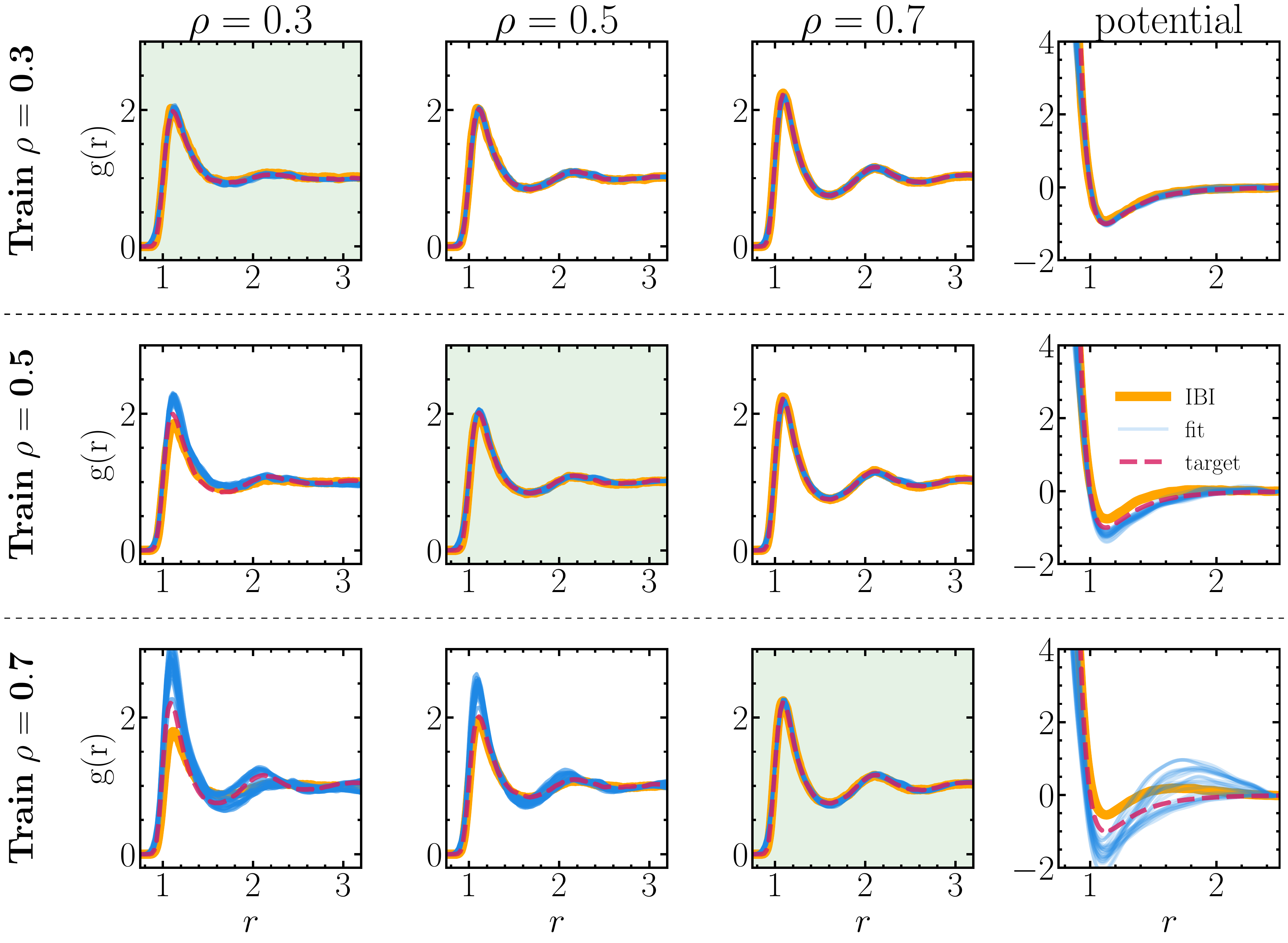}
    \caption{Transferability of DiffSim and IBI potentials fitted to reproduce $g(r)$ of Lennard-Jones potentials at different densities. Each row contains the results for the simulated g(r) at three densities (0.3, 0.5 and 0.7 from left to right) with potential fitted at a particular density (light green background, 0.3, 0.5 and 0.7 top to bottom) The last column reports the learned potentials compared the ground truth.
    We include learned pair potentials obtained from 10 independent runs. }
    \label{fig:lj_trans}
\end{figure}

As a first example, we apply our method to simple Lennard-Jones(LJ) systems at different densities. We obtain target RDFs for LJ particles simulated at three densities $\rho=(0.3, 0.5, 0.7)$ with temperature set at $1.0$, and we use the obtained RDFs as the optimization target for our DiffSim protocol. We obtain the target RDFs by simulating the systems for 500,000 steps after the initial equilibration of 100,000 steps. We use a time step of 0.005. To test the transferability of learned potentials at different system densities, we first learn pair potentials at a single density and use the learned potential to simulate the system with the other two densities that are not used for learning. The learning results are presented in \cref{fig:lj_trans}, which also includes a comparison with the potentials obtained using IBI. Our method shows an accurate fitting of the RDFs for the dedicated densities, but the level of transferability varies. The potential learned at the low density of $\rho=0.3$ demonstrates excellent transferability between different densities, and the learned potential also agrees well with the ground-truth LJ potential. As the density of the system used for fitting increases, the learned potentials show high variability between replicate experiments, with overall poorer transferability to systems that are not used for training. Many of the learned potentials are dissimilar to the ground-truth potential, featuring concave-down shapes. For a high-density system with $\rho=0.7$, neither IBI nor DiffSim produces potentials that resemble the ground-truth LJ potential. The learned potentials for $\rho=0.7$ also do not transfer well to $\rho=0.7, 0.3$ with varied magnitudes of deviation. 

\begin{figure}
    \centering
    \includegraphics[width=0.45\textwidth]{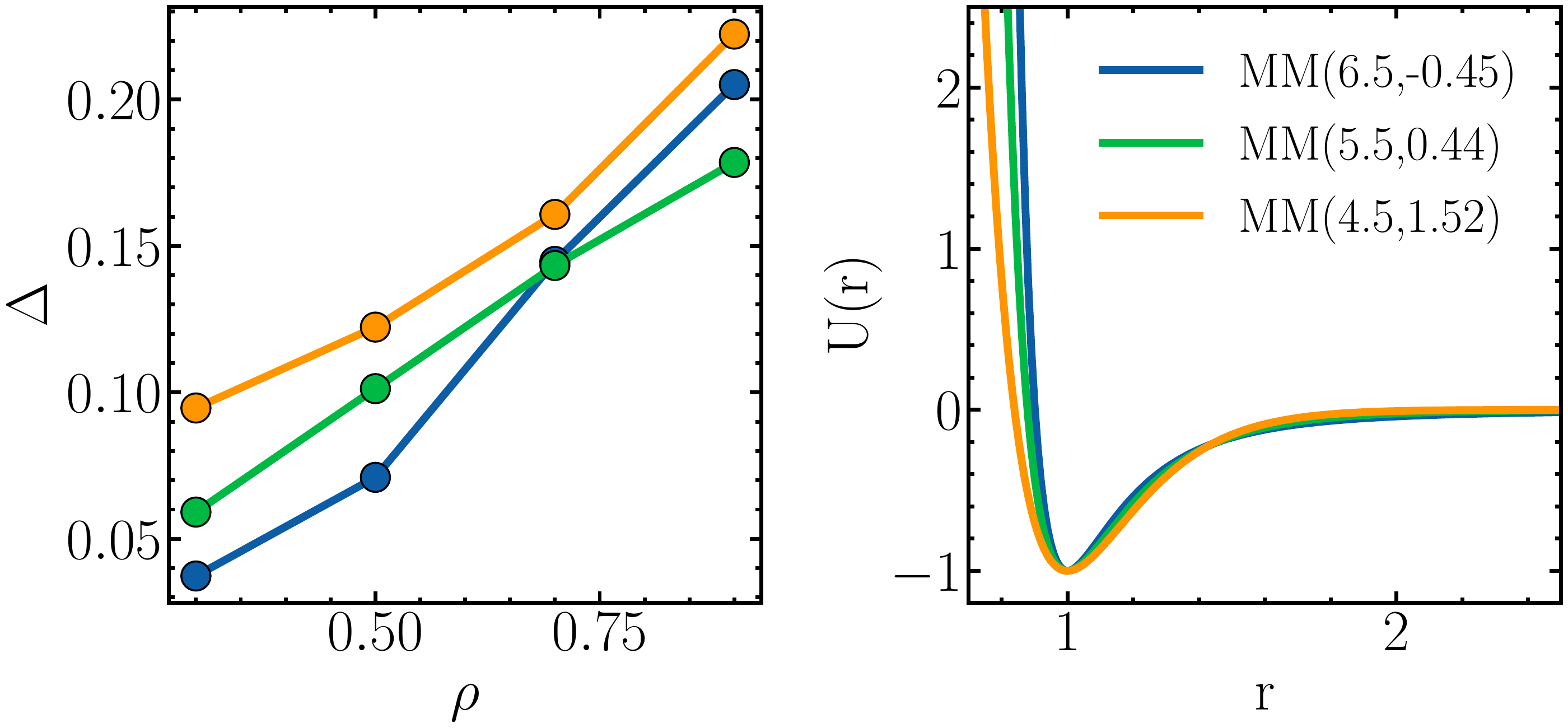}
    \caption{Left: Calculated mean deviation ($\Delta$) between the fitted potentials and the ground-truth modified Morse potentials. For the three potentials tested, the learned potentials manifest less deviation as density decreases. Right: Modified Morse potentials used for learning.}
    \label{fig:morse_dev}
\end{figure}

To demonstrate that the observed trend is not specific to the LJ potential, we tested our methods on the Modified Morse (MM) Potentials with different softness at different densities \cite{cheng2007modified}. The MM potential takes the following form: 

    \begin{equation}
        MM(\rho, \psi)(r) = \frac{e^{2\rho(1 - r^{\psi})} - 2 e^{\rho(1 - r^{\psi}} - A)}{1 + A}
    \end{equation}

where $A = 0 $ for $\psi \leq 0$ and $A = e^{2 \rho / \psi} - 2 e^{\rho / \psi}$ for $\psi < 0$. One can tune $(\rho, \psi)$ to generate a diverse set of pair potentials. We tested our method for three different potentials $MM(6.5, -0.45)$, $MM(5.5, 0.44)$, and $MM(4.5, 1.52)$ at four different densities $\rho=(0.3, 0.5, 0.7, 0.9)$ to infer pair potentials from target RDFs in those systems. For quantitative evaluation, we calculate the average deviation of learned potentials from the ground truth using the metric:

\begin{equation}
    \Delta = \int_{r_0}^{r_c} |U_{fit}(r) - U_{true}(r)| dr
\end{equation}

where we choose $r_c=2.5$ and $r_0=0.9$. \cref{fig:morse_dev} shows that the learned potentials deviate less from the ground truth as the density decreases, which is consistent with the trend observed for LJ systems. This is also in line with observations in the past literature.\cite{moore2014derivation} For denser systems, we attribute the low sensitivities of the learned potentials to dense local packing that downplays the effect of intermolecular forces, largely due to the ``entropic effect of packing". \cite{chandler1983van, wang2020sensitivity}


\begin{figure}
    \centering
    \includegraphics[width=0.45\textwidth]{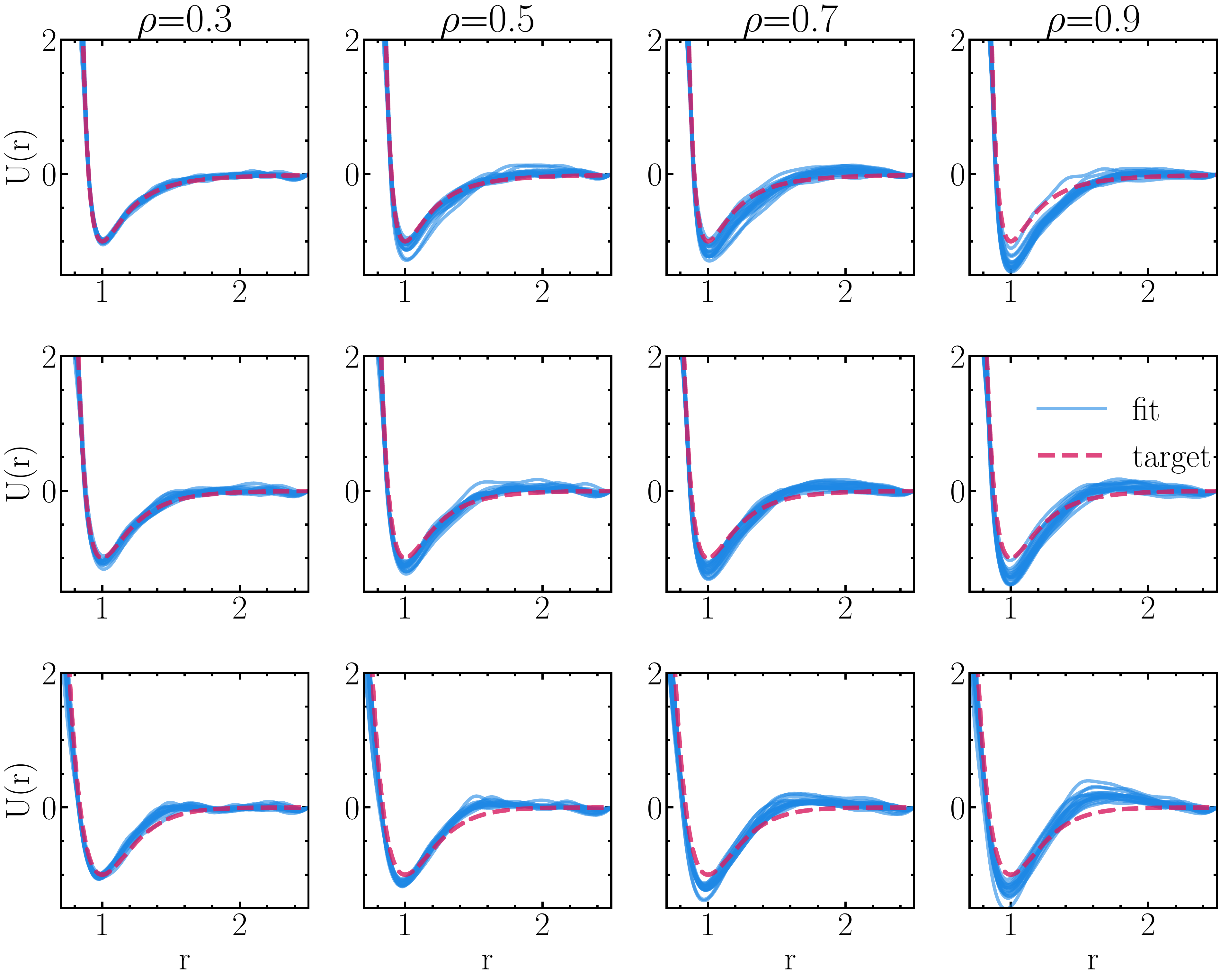}
    \caption{Comparison of fitted potentials with modified Morse potentials $\textbf{MM(6.5, -0.45)}$ (top), $\textbf{MM(5.5, 0.44)}$ (middle), and $\textbf{MM(4.5, 1.52)}$ (bottom). The figure includes learned pair potentials obtained from 10 independent runs. }
    \label{fig:morse}
\end{figure}

\subsection{Learning Pair Potentials for Binary Systems}

Our DiffSim-based learning protocol is highly flexible. For example, it can be modified to incorporate multiple learning targets to learn pair interactions between different particle types. During training, this involves the simultaneous optimization of different interaction functions between different types of particles. 

To demonstrate the efficacy of our method in this case, we construct a similar example as presented in Rosenberger \textit{et al.} \cite{rosenberger2019relative}. The binary mixture system involves three types of LJ interactions with the following parameters: $\{ (\epsilon_{AA}: 1.0, \sigma_{AA}: 0.9), (\epsilon_{AB}: 1.0, \sigma_{AB}: 1.0), (\epsilon_{BB}: 1.0, \sigma_{BB}: 1.1) \}$.  We simulate 500,000 steps at a temperature of $k_B T =1.0 $ with a time step of 0.005 to generate the target RDFs for $x = (0.25, 0.5, 0.75)$, where $x$ is the composition of particle A: $\frac{n_A}{n_A + n_B}$. We optimize based on targets at $x=0.25$ and test the transferability of the learned potential at $x=(0.5, 0.75)$. During the simulation, we learn the three types of interaction simultaneously by optimizing a joint loss of equally weighted loss terms. 

    \begin{equation}
    \begin{aligned}
      L &=  L_{AA} + L_{AB} + L_{BB} \\
        &=  \int_r (g_{\simulation}^{AA}(r) - g_{\reference}^{AA}(r))^2  dr + \int_r (g_{\simulation}^{AB}(r) - g_{\reference}^{AB}(r))^2  dr \\ &+  \int_r (g_{\simulation}^{BB}(r) - g_{\reference}^{BB}(r))^2  dr 
     \end{aligned}
     \label{eq:multi_loss}
    \end{equation}

\begin{figure}
    \centering
    \includegraphics[width=0.4\textwidth]{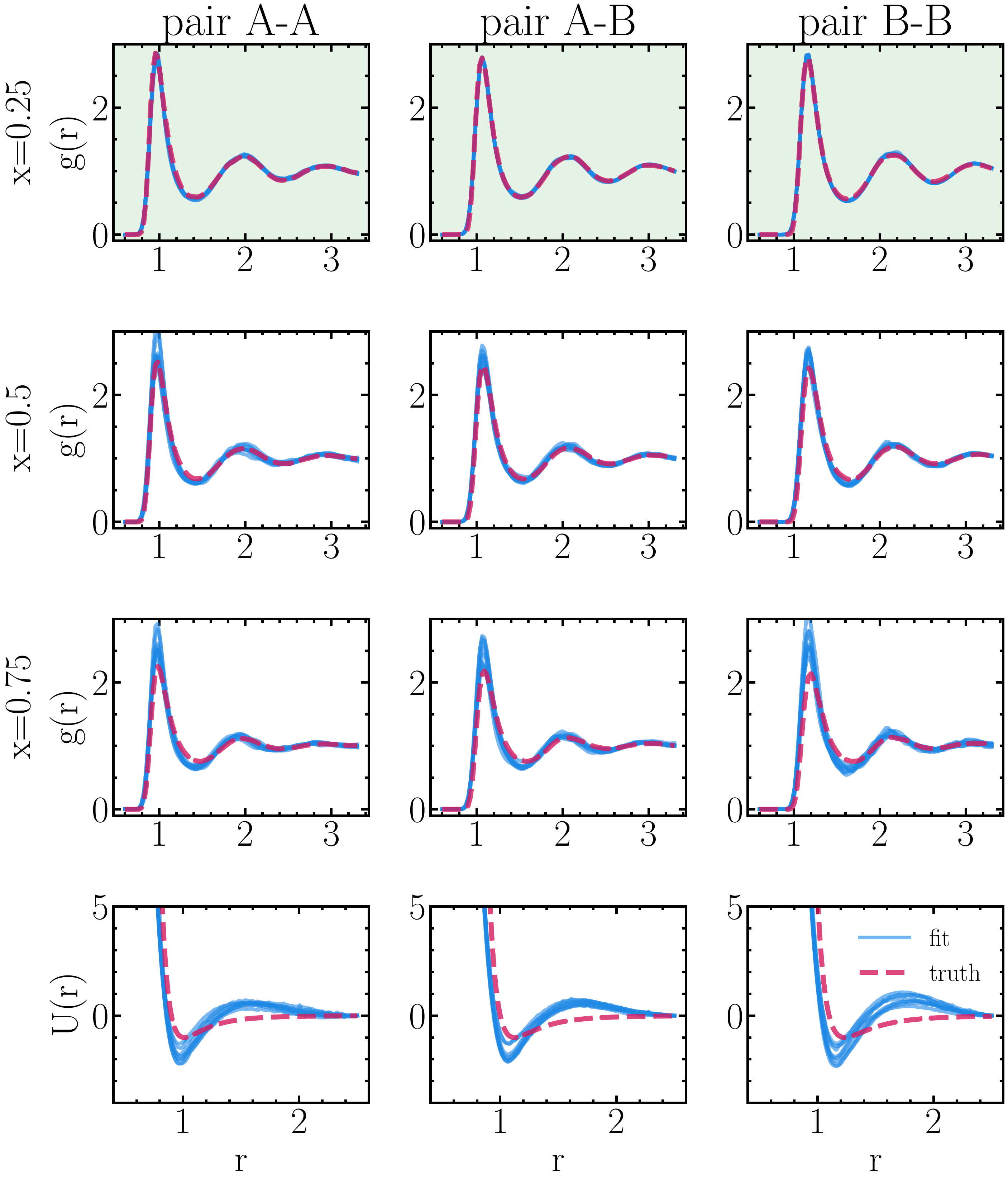}
    \caption{We apply the pair potentials learned at $x=0.25$ (highlighted with light green background) to systems with $x=(0.5, 0.75)$. Unlike ground truth potentials of the LJ type, the learned potentials feature concave-down shapes.}
    \label{fig:binary_0.25}
\end{figure}

We optimize three potentials for 300 epochs using the Adam optimizer with a learning rate of 0.001, with each epoch consisting of 250 MD steps. In \cref{fig:binary_0.25}, we show that DiffSim accurately reproduces the three target RDFs for $x=0.25$. However, as the learned potential correctly recovers the repulsive part of the target potential, the attractive part shapes are dissimilar to the ground truth and feature a concave-down shape. As a transferability test, we apply the learned potential at $x=0.25$ to systems with different compositions, that is, $x=0.5$ and $x=0.75$, and observe poor transferability. Specifically, the learned potentials overestimate the height of the first peak for the target RDFs at $x=0.5$ and $x=0.75$. In this case, the learned potential overfits to the composition state point on which it is trained. 

\begin{figure}
    \centering
    \includegraphics[width=0.4\textwidth]{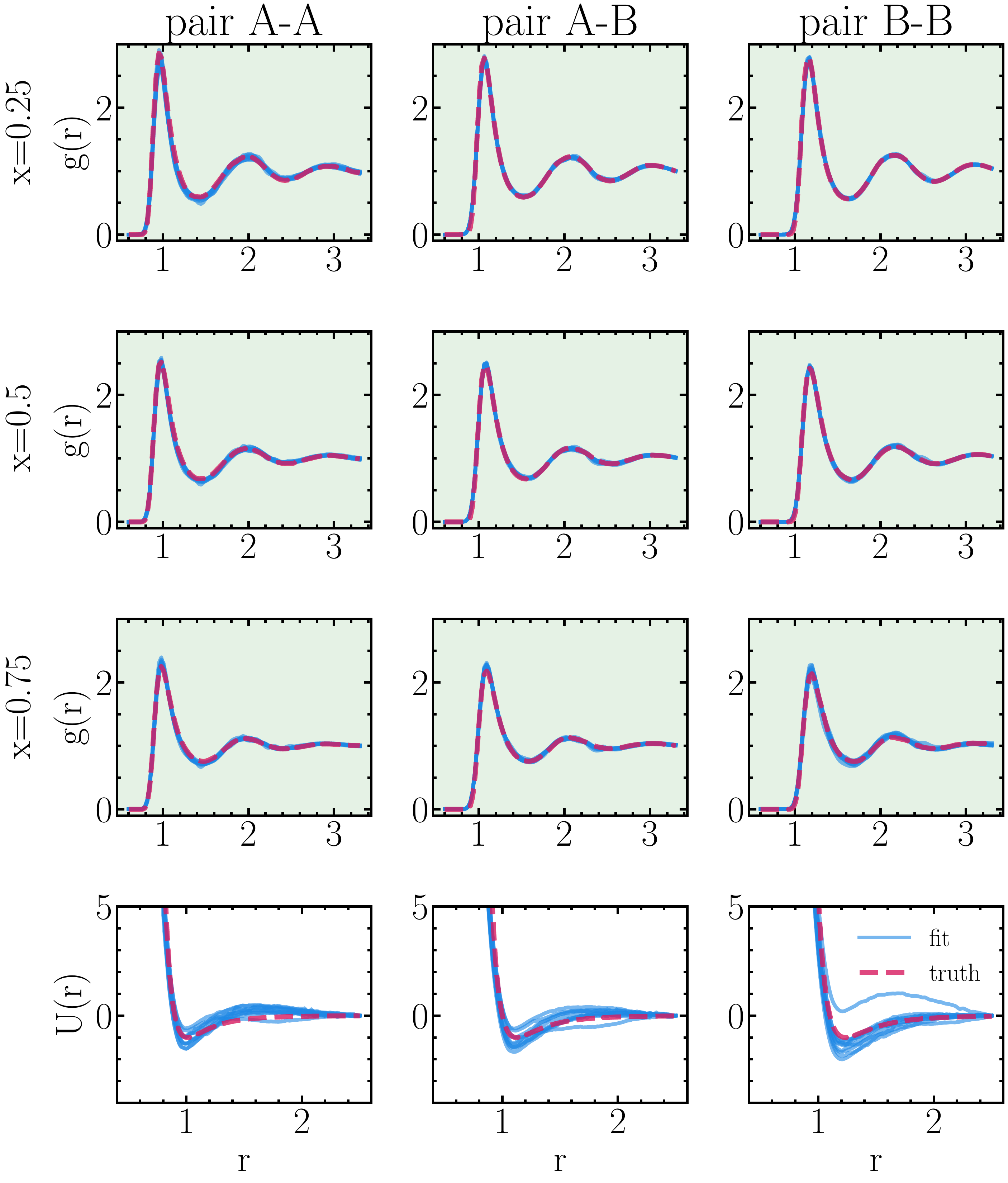}
    \caption{RDFs and pair potentials obtained from simultaneous learning of pair potentials for systems at $x=0.25, x=0.5, x=0.75$. The learned potentials resemble the ground truth potentials. The learned potentials accurately recover the RDFs at all compositions.}
    \label{fig:binary_0.25_0.5_0.75}
\end{figure}

For improved transferability, our model can also be configured to train systems simultaneously in different states. This involves applying DiffSim to simulate multiple systems and combining multiple gradient updates to the potential arising from the different trajectories. In practice, we initialize all three systems and simulate them in parallel with the same shared potentials. Gradient updates are collected through \cref{eq:grad} and \cref{eq:loss} from each trajectory, and the pair potentials of AA, AB, and BB are updated with the combined three gradients from the three state points. For each training epoch, three systems are simulated with gradients accumulated on the parameters of the pair potential with the total loss as the sum of individual losses evaluated for each system, i.e., $L_{tot} = L_{x=0.25} + L_{x=0.5} + L_{x=0.75}$. Each loss takes the form described in \cref{eq:multi_loss}, giving us 9 optimization targets in total, namely the RDF of AA, AB, and BB at the three concentrations. Our training results in \cref{fig:binary_0.25_0.5_0.75} show that potentials can be learned to simultaneously fit target RDFs at multiple systems and state points at the same time. Compared to potentials solely fitted at $x=0.25$ (\cref{fig:binary_0.25}), the learned potentials in all three states show a better resemblance to ground-truth LJ potentials for both repulsive and attractive portions.

\subsection{Learning Transferable Coarse-Grained Potentials}

\begin{figure*}[htb]
    \centering
    \includegraphics[width=0.65\textwidth]{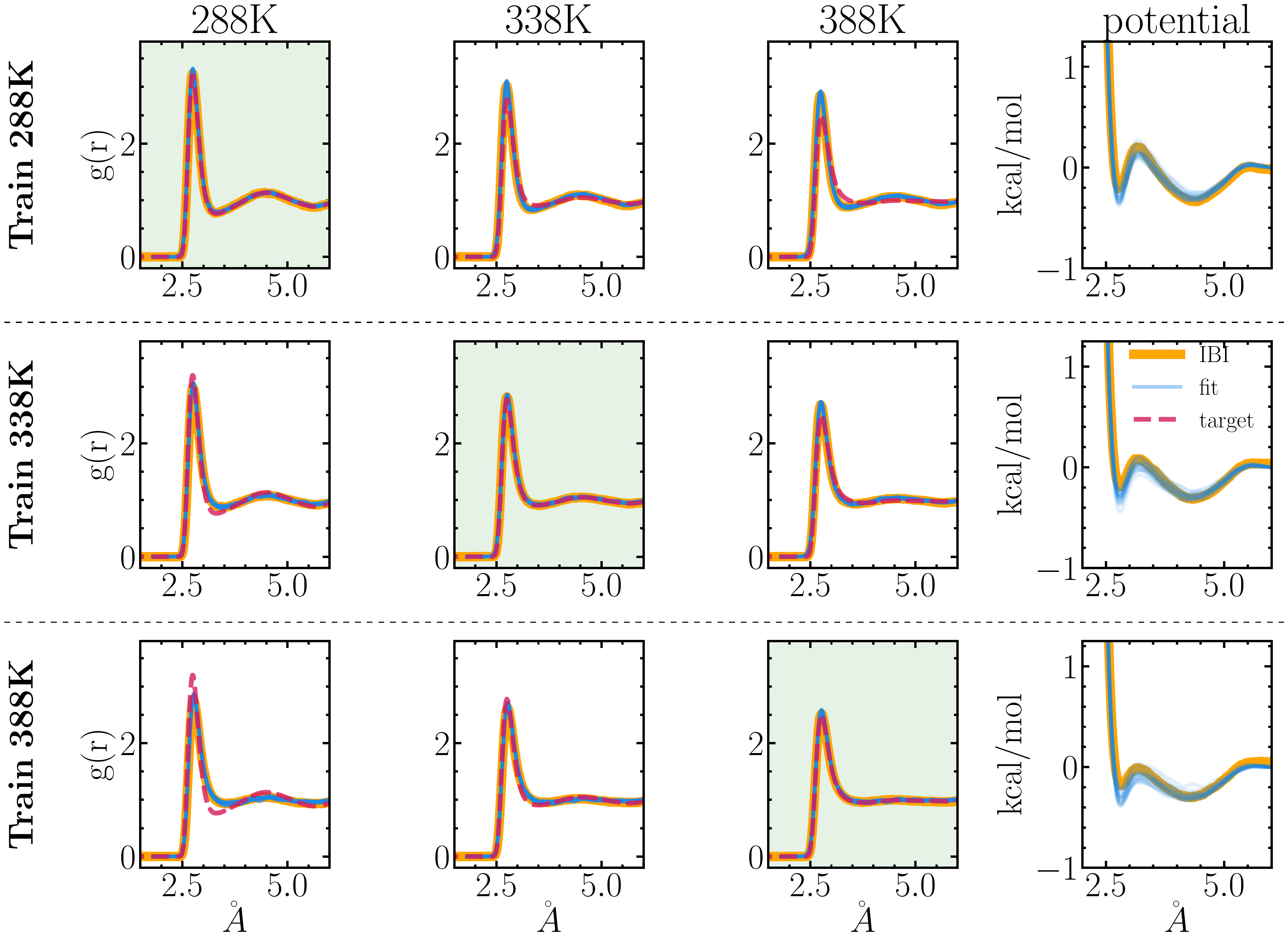}
    \caption{Learning coarse-grained water potential at single state points and transfer to other state points. The results include learned pair potentials and simulated RDFs from 20 independent runs with DiffSim. The learned potentials using IBI is also included. }
    \label{fig:water_trans}
\end{figure*}

As the third example, we show that DiffSim can learn pair potentials for CG simulations of water with a temperature-transferable functional form. For the ground truth simulations, we simulate all-atom water molecules at $1g/cm^3$ at 288K, 338K and 388K with the SPC/E force field.\cite{berendsen_missing_1987} We choose to map each water molecule onto the oxygen atom and use the oxygen-oxygen RDF as our fitting target. We first run a single-state fitting experiment at single temperatures and use the learned potentials to simulate systems at other temperatures. For each system, we train the potential using DiffSim for 500 epochs, with each epoch including 180 simulation steps with a time step of 2fs. For comparison, we also performed the IBI at all three temperatures.  \cref{fig:water_trans} shows the fitting and transferability results. As DiffSim and IBI manage to accurately fit single states, they manifest large RDF errors when using similar systems at other temperatures. The learned potentials using IBI and DiffSim show similar shapes, with an inner potential well at $2.7 \angstrom$, approximately where the RDF of water has the first peak. Furthermore, the depth decreases with temperature, showing a clear temperature dependence of the pair potential. This observation agrees with the water potentials obtained in the past literature \cite{johnson2007representability, pretti2021microcanonical}, suggesting that an explicit temperature dependence is required for the pair potential to be transferable.

To incorporate the temperature-dependent nature of CG potentials, we made our neural pair potentials temperature-dependent for improved transferability across temperatures. With AD, our method enables more flexible forms of pair potentials. Specifically, we optimize pair potentials with explicit temperature dependence. Inspired by previous work studying the energy and entropy decomposition of CG potentials,\cite{kidder2021energetic, pretti2021microcanonical} we propose the following form of neural pair potential.

    \begin{equation}
        u_T(r, T) = u_1(r) - k_B T u_2(r)
    \end{equation}

where $u_1$ and $u_2$ are two distinct neural pair potentials. $u_1(r)$ can be understood as the energy contribution to pair potentials, and $u_2(r)$ is the entropy contribution. For training, we initialize three coarse-grained water systems and simulate them at different temperatures. We simulate and train the three systems simultaneously to optimize $u_1(r)$ and $u_2(r)$ jointly. In \cref{fig:water_trans}, we show the RDFs obtained and the learned pair potentials at the three different temperatures used for learning. The learned potentials at different temperatures show similar shapes. All three potentials have a characteristic barrier around 3.2 $\AA$, and the change in temperature affects their height. Compared to the potential obtained by fitting on single states, the learned temperature-dependent pair potential shows excellent transferability to all three temperatures used for optimization. To quantify the deviation between the simulated RDF and the target RDF, we use the following metric proposed by Pretti. \textit{et al.}: 

\begin{equation}
    D = 4 \pi \rho \int_0^{r_c} r^2 (g_(r) - g_{\reference}(r))^2 dr
\end{equation}

\begin{figure}[htb]
    \centering
    \includegraphics[width=0.48\textwidth]{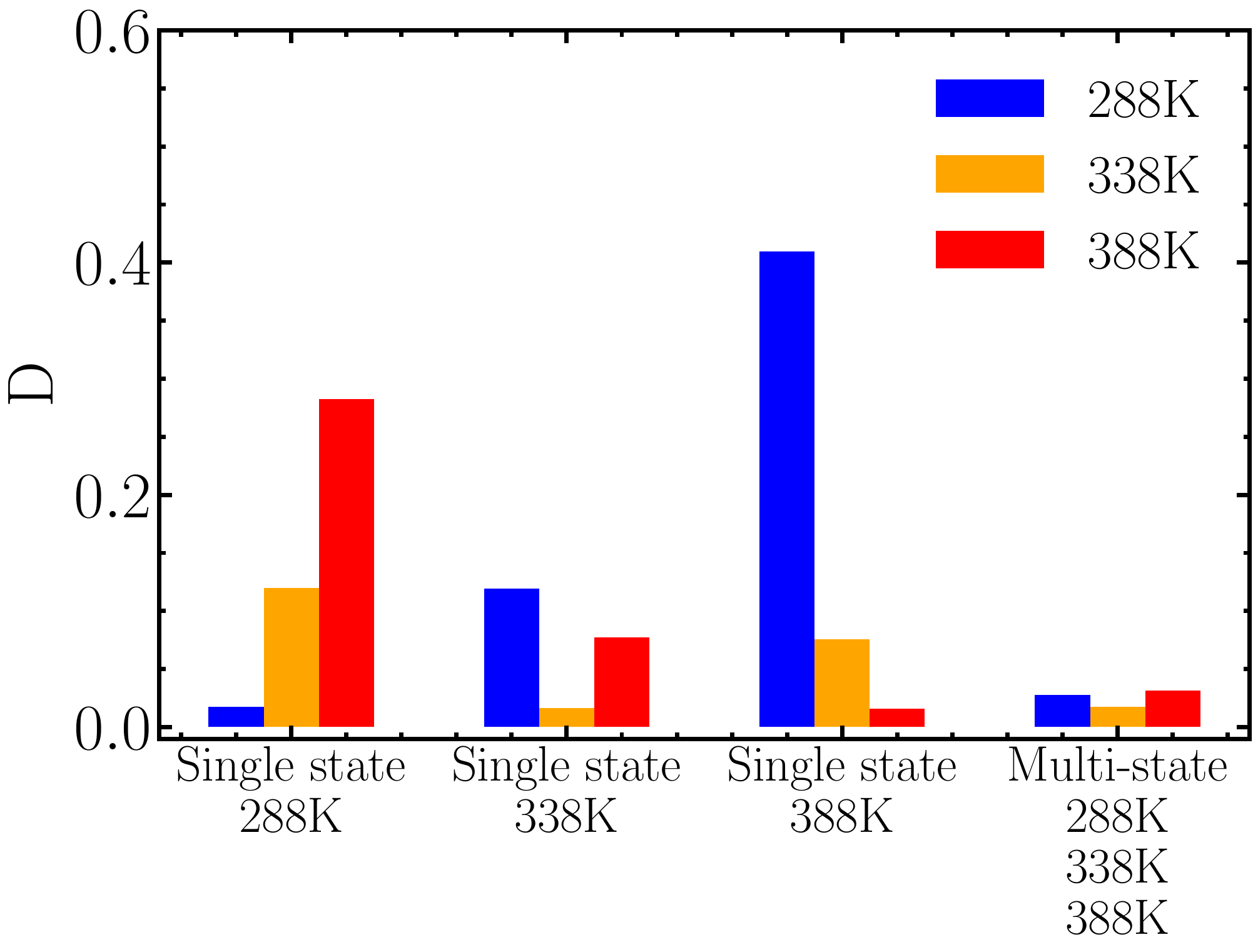}
    \caption{Quantified RDF error ($D$) of learned potentials simulated at different state points (288K, 338K, 388K).}
    \label{fig:water_err}
\end{figure}

In \cref{fig:water_err}, we compare the deviation of the simulated RDF obtained from different training protocols. The result shows that fitting in a single state can produce an accurate RDF in the state used for fitting, but it does not transfer well to other states. The temperature-transferable potential produces overall lower RDF deviations for all three states. 
    
\begin{figure*}[htb]
    \centering
    \includegraphics[width=0.65\textwidth]{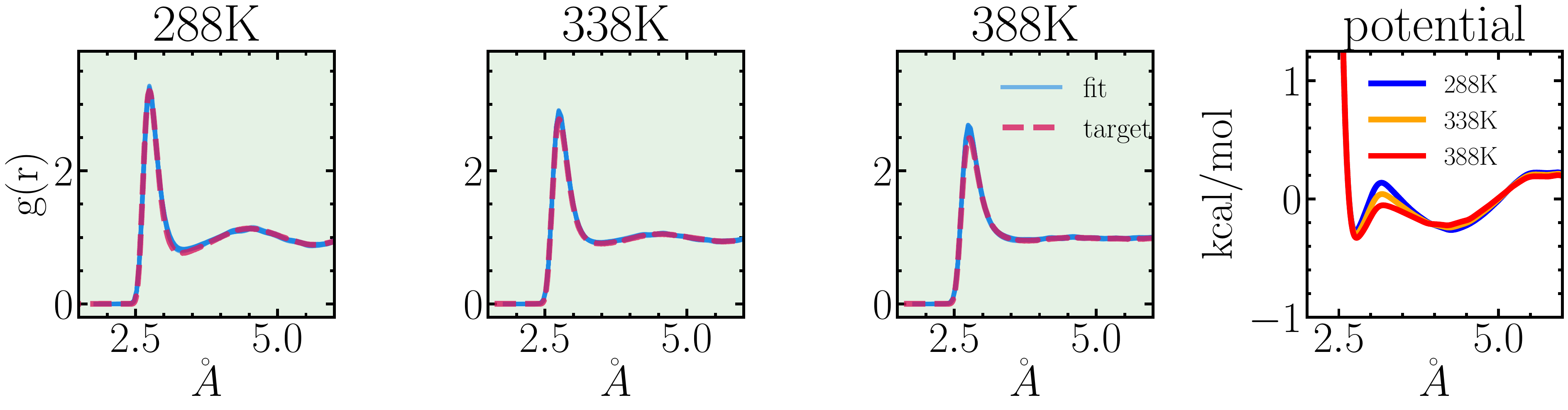}
    \caption{Simulated RDF and learned transferable potential at 288K, 338K, 388K. The learned potential shows temperature-dependent barrier height, with the height increasing with lowering in temperature.}
    \label{fig:water_temp}
\end{figure*}

\section{Conclusion}
In this work, we propose a flexible method based on DiffSim to directly learn pair potentials. We demonstrate our methods in several computational examples. First, we apply our method to learning simple pair potentials where the ground truth is known. We show that our method recovers a diverse set of possible pair potentials given a target RDF and, therefore, can be used to probe sensitivities for fitting pair potentials. For a dense system, our results are in line with previous reports, indicating that fitted potentials can be highly insensitive. \cite{wang2020sensitivity, potestio2013henderson} In contrast, the fitting for the dilute system shows a better sensitivity, with the obtained potentials in reasonable agreement with the ground truth potential. We further demonstrate the multistate fitting capabilities of the method by applying it to binary mixtures that involve learning of multiple pair interactions under multiple composition conditions. Learning involves the incorporation of multiple loss functions in different RDF targets. Our experiment shows that simultaneous learning on systems with different compositions is possible and produces potentials that have better transferability. Our method is also applied to learning the CG pair potentials. We demonstrate the learning of water CG potentials at three different state points using a single temperature-dependent CG potential that is more transferable than a potential obtained from fitting a single state point.

The proposed DiffSim protocol can be understood as a gradient-based extension of the iterative Boltzmann inversion. Our method directly optimizes potentials with the gradient signal obtained from simulations. It allows simultaneous simulation and fitting for systems at different temperatures and compositions, allowing direct parameterization of pair potentials to improve transferability. The learning pipeline can be flexibly set up with the incorporation of multiple optimization objectives and allows for the optimization of a broader range of pair potentials, such as pair potentials with explicit thermodynamic dependence. Our method opens up broader possibilities for learning-based simulations: objectives can be extended to include more observables such as stress tensors and velocity autocorrelation functions to enable gradient-based top-down parameterization of force fields.

DiffSim learns potentials by directly optimizing unrolled simulation operations that involve updates of positions, velocities, and forces. The procedure directly learns to match an observation without atomistic data, which requires extensive sampling that often requires active learning. Compared to force-based optimization of machine learning potentials, our method produces direct simulation feedback and guarantees the stable production of desired observables. In comparison, relying on matching atomistic forces, highly flexible machine learning force fields often suffer from unstable dynamics, which is common even for models showing very low force errors.\cite{stocker2022robust} In comparison, pair potentials constrain interactions between pairs of particles and are therefore easier to interpret and more efficient to simulate than machine learning potentials. Although observable-informed and gradient-based, our learning protocol can also be combined with force matching to incorporate top-down and bottom-up parameterization into optimization.

In this work, we present a flexible machine learning method that uses differentiable simulations to directly parameterize pair potentials from radial distribution functions. We anticipate that the proposed framework provides improved modeling flexibility for multiscale simulation of molecular liquids.  

\section*{Acknowledgment}
This work is supported by Toyota Research Institute.

\section*{Author Declaration}

\subsection*{Conflict of Interest}
The authors have no conflicts to disclose.

\subsection*{Author Contributions}
\textbf{Wujie Wang}: Conceptualization (lead); Investigation (equal); Writing – original draft (equal); Writing – review \& editing (equal). 
\textbf{Zhenghao Wu}: Conceptualization (supporting); Investigation (equal); Writing – original draft (equal); Writing – review \& editing (equal).
\textbf{Rafael Gómez-Bombarelli}: Conceptualization (supporting); Supervision; Funding acquisition; Writing – review \& editing (equal).

\section*{Data Availability}
The data and code that support the findings of this study are available in \href{https://github.com/torchmd/mdgrad}{https://github.com/torchmd/mdgrad}.

\section{Appendix}

\subsection{Model and training hyper-parameters}

Here, we detail the hyperparameters used in our experiments. For all training runs, a learning rate scheduler is used to dynamically decrease the learning when the pre-defined convergence criterion is met. The learning rate is expected to decrease by half when the improvement in training loss is less than $0.1 \% $.

\textbf{Lennard Jones and MM systems.} The model is trained for 500 epochs, with each epoch comprising 120 simulation steps. The learning rate is set at $0.002$.  Neural pair potentials are constructed with Gaussian-smeared distance inputs of 100 bins ($K = 100$) with $d=0.1$. The smeared inputs are fed into a multilayer perceptron of 3 hidden layers with a hidden layer size of 128. The activation function that we use is ELU.\cite{clevert2015fast} The prior potential $u_{prior}(r)$ that we use takes the form $ \epsilon (\frac{r}{\sigma})^{10}$ with $\sigma = 0.9$ and $\epsilon = 0.4$. During simulations, the neighbor list is updated for every integration step.

\textbf{Binary mixtures.} The model is trained for 500 epochs, with each epoch comprising 120 simulation steps. The learning rate is set at $0.0003$. Three neural pair potentials are initialized to represent the three types of pair interaction involved in the system. For the computation of RDF loss, each pair statistics is equally weighted. Each pair potential uses 100 bins ($K=100$) to smear all distance inputs with $d = 0.15$. The parameterization MLP consists of 3 hidden layers with each layer of 128 neurons. The activation function used is SELU. \cite{klambauer2017self}  The prior potential $u_{prior}(r)$ we use the form $ \epsilon (\frac{r}{\sigma})^{6}$ with $\sigma = 1.0$ and $\epsilon = 2.0$. Before training, a pre-training step is performed to initialize neural network potentials with shapes similar to the Boltzmann inverted potentials derived from the target RDF. This step is used to facilitate faster convergence of training and sampling of relevant parts of the configuration space. 

\textbf{Temperature-transferable water simulations. } The model is trained for 500 epochs, with each epoch comprising 190 simulation steps. The learning rate is set at $0.00065$. We similarly apply 3-layer neural networks to represent $u_1$ and $u_2$ with hidden layers of 115 neurons. The pair distances are smeared with $K=120$ equally spaced Gaussian kernels with $d=0.15$. The activation function used is ELU. \cite{clevert2015fast}.  $u_{prior}$ is used with $\sigma=40$ kcal/mol and $\epsilon=1.7 \angstrom$. A pre-training procedure is also performed using the Boltzmann inverted RDF at $298K$ as the target.

\subsection{Supplementary results}

\begin{figure}
    \centering
    \includegraphics[width=0.5\textwidth]{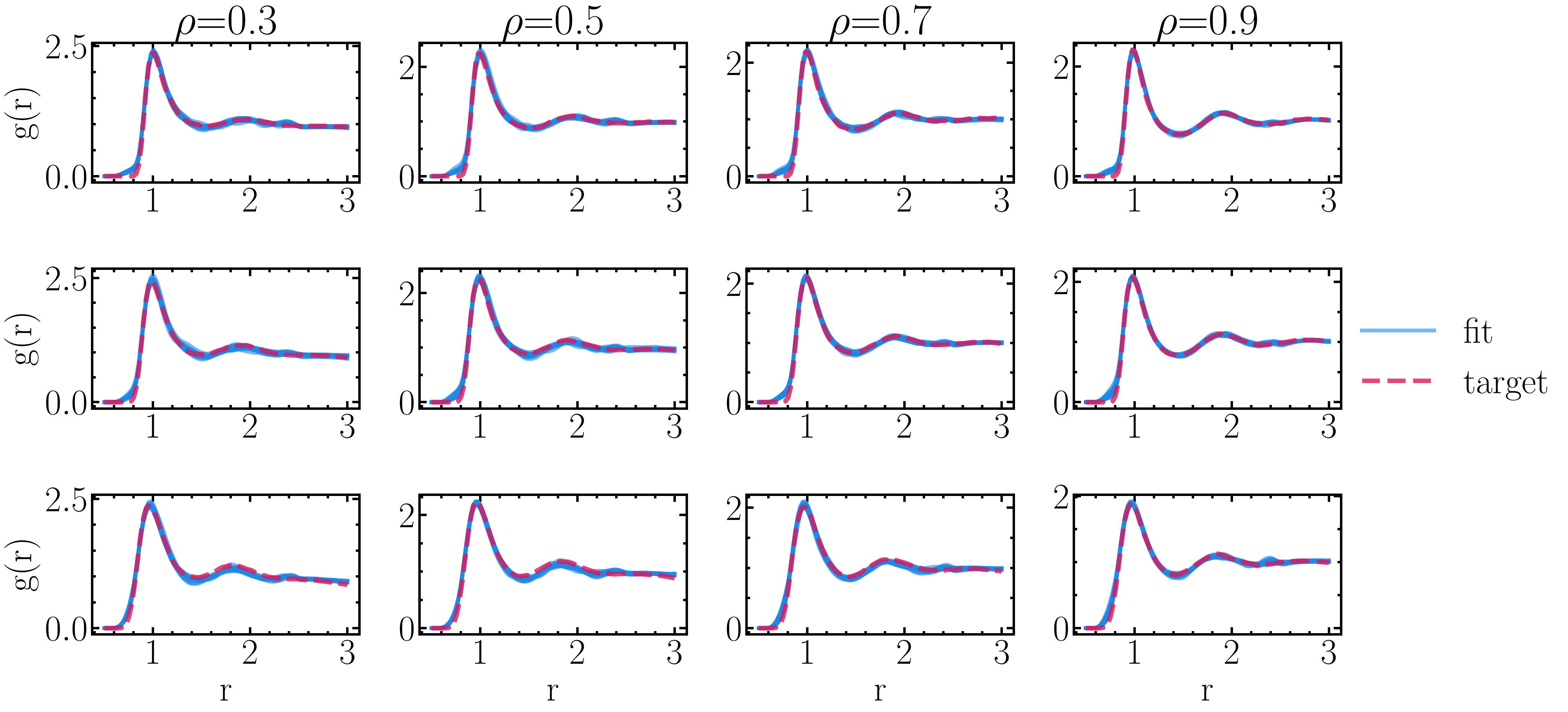}
    \caption{Comparison between ground-truth RDFs and RDFs from simulations using learned pair potentials for different MM potentials at different densities. From top to bottom rows, the figures correspond to RDFs generated by learned potentials for MM(6.5, -0.45) , MM(5.5, 0.44), and MM(4.5, 1.52) systems respectively.}
    \label{fig:morse_rdf}
\end{figure}

We provide additional results in \cref{fig:morse_rdf} to show the RDFs generated by learned pair potentials for different MM systems at different densities. \cref{fig:morse_rdf} provides supplmentary information for \cref{fig:morse}. The simulations are carried out for 100,000 steps with each step being 0.005. 

\subsection{Molecular dynamics simulations of water}
The FG model of water used in this work is an all-atom water model described by SPC/E force field\cite{berendsen_missing_1987}. The large-scale atomic / molecular massively parallel simulator (LAMMPS)\cite{plimpton_fast_nodate,thompson2022lammps} is used to perform MD simulations of systems with $N=1500$ water molecules at constant density $\rho=1.0$ $\mathrm{g/cm^3}$ and at temperatures of $T=288K$, $T=338K$ and $T=388K$, with a time unit $\delta t=1$ fs. We apply periodic boundary conditions in the x, y and z directions, and the No\'{s}e-Hoover thermostat\cite{nose_unified_1984,hoover_canonical_1985} with a damping time of $\tau=100$ fs to keep the temperature during the simulation. The water systems are first equilibrated for $10$ ns. Subsequently, it was subjected to a simulation run of $1$ ns to collect data every 0.1 ps to calculate the reference radial distribution functions of oxygen atoms with bin size 0.001 nm. In the CG representation, a water molecule is replaced by a single bead placed on the oxygen atom.  CG simulations are also performed in LAMMPS. \cite{plimpton_fast_nodate,thompson2022lammps}

\subsection{Protocol for Iterative Boltzmann Inversion}

\textbf{Lennard-Jones systems.} The RDFs of the MD simulations with LJ potentials are the target distribution fed to the IBI machinery. The cutoff radius used for IBI optimization and simulations with IBI refined potential is $r_{cut}=2.0$ with space size $\delta r=0.01$ $\sigma$. For each state point, the MD simulation performed under the canonical ensemble in each IBI iteration consisted of $5\times 10^5$ equilibration steps and $5\times 10^5$ sampling steps for the calculation of the RDF. The entire IBI optimization takes between 10 and 20 iterations to yield converged distributions that match the target RDFs.

\textbf{Water simulations.} The RDFs from the SPC/E water simulation are obtained by mapping the CG beads mapped to the oxygen atoms and are fed into the IBI optimization protocol. The cutoff radii used for IBI optimization and CG simulations are $r_{cut}=0.625$ nm with a space size $\delta r=0.0005$ nm. For each state point, the CG-MD simulation performed in the canonical ensemble in each IBI iteration consisted of $2\times 10^6$ equilibration steps and $1\times 10^6$ sampling steps for the calculation of the RDF. The entire IBI optimization takes between 10 and 20 iterations to yield converged distributions that match the target RDFs.

\bibliography{aipsamp}

\end{document}